\documentclass{PoS}

\usepackage{graphicx}
\usepackage{epsfig}
\usepackage{dcolumn}
\usepackage{bm}
\usepackage{cite}

\def\al{\alpha}
\def\be{\beta}
\def\ga{\gamma}
\def\De{\Delta}
\def\de{\delta}

\def\W{\Omega}
\def\w{\omega}

\def\cpt{$\chi$PT }

\def\3d{3-D}

\newcommand{\dd}{\ensuremath{\mathrm{d}}}
\newcommand{\hq}{\hspace*{0.5em}}

\title{Nucleon Spin Polarisabilities from Polarised Deuteron Compton
  Scattering
}

\ShortTitle{Polarised $\gamma$ d Scattering}

\author{\speaker{Harald W. Grie{\ss}hammer}\\
  Center for Nuclear Studies, The George Washington University, Washington, DC 20052\\
  E-mail: \email{hgrie@gwu.edu}}

      \author{Deepshikha Shukla\\
        Center for Nuclear Studies, The George Washington University, Washington, DC 20052\\
        E-mail: \email{dshukla@email.unc.edu}}

      \abstract{We present recent results on elastic deuteron Compton
        scattering calculations for polarised beans and targets up to
        next-to-leading order within Chiral Effective Field Theory in the
        Small Scale Expansion variant to implement a dynamical $\Delta(1232)$
        degree of freedom. A simple power-counting argument discloses that
        np-intermediate rescattering states must be explicitly included at
        leading order already. This automatically results in the correct
        Thomson limit and guarantees current conservation. In view of ongoing
        effort at MAXlab, proposals at HI$\gamma$S and plans at MAMI, we
        address in detail single- and double-polarised observables with
        linearly or circularly polarised photons on both unpolarised and
        vector-polarised deuterons. Our results indicate that several of the
        polarisation observables can be instrumental to extract not only
        spin-independent nucleon polarisabilities, but also the so-far
        practically un-determined spin-dependent polarisabilities which
        parameterise the stiffness of the nucleon spin in external
        electro-magnetic fields.  Amongst the questions addressed are:
        convergence of the expansion for including the $\Delta$, the r\^ole of
        the np-rescattering contributions, and sensitivity to the deuteron
        wave function. An interactive \emph{Mathematica 7.0} notebook of these
        findings is available from hgrie@gwu.edu.}

\FullConference{6th International Workshop on Chiral Dynamics, CD09\\
        July 6-10, 2009\\
        Bern, Switzerland}

\begin{document}

\section{Introduction}

As the nucleon is not a point-like target, the photon field displaces its
charged constituents, inducing a non-vanishing multipole-moment. Low-energy
Compton scattering $\gamma N\to\gamma N$ of real photons probes therefore the
temporal response of the low-energy degrees of freedom inside the nucleon,
encoded in the nucleon polarisabilities~\cite{barry}. These are parameterised
by the most general interaction between a nucleon $N$ with spin
$\vec{\sigma}/2$ and an electromagnetic field of non-zero energy
$\omega$:
\begin{eqnarray}
  \label{polsfromints}
  \lefteqn{\mathcal{L}_\mathrm{pol}=2\pi N^\dagger\big[{\alpha_{E1}(\omega)}\vec{E}^2+
    {\beta_{M1}(\omega)}\vec{B}^2+{\gamma_{E1E1}(\omega)}
    \vec{\sigma}\cdot(\vec{E}\times\dot{\vec{E}})}\\
  &&\!\!\!\!\!\!\!+{\gamma_{M1M1}(\omega)}
  \vec{\sigma}\cdot(\vec{B}\times\dot{\vec{B}})
  -2{\gamma_{M1E2}(\omega)}\sigma_iB_jE_{ij}+
  2{\gamma_{E1M2}(\omega)}\sigma_iE_jB_{ij}+\dots \big]N\nonumber
\end{eqnarray} 
Here, the electric or magnetic (${X,Y=E,M}$) photon undergoes a transition
${Xl\to Yl^\prime}$ of definite multipolarity ${l,l^\prime=l\pm\{0,1\}}$;
${T_{ij}:=\frac{1}{2} (\de_iT_j + \de_jT_i)}$. Its coefficients are the
\emph{energy-dependent} or \emph{dynamical polarisabilities} of the
nucleon~\cite{Hi04}. Most prominently, there are six dipole-polarisabilities:
Two spin-independent ones parameterise electric and magnetic
dipole-transitions, $\alpha_{E1}(\omega)$ and $\beta_{M1}(\omega)$, whose
static values $\bar{\alpha}\equiv\alpha_{E1}(\omega=0)$ and
$\bar{\beta}\equiv\beta_{M1}(\omega=0)$ are often simply called ``the
polarisabilities''.  For the proton, all extractions agree within their
uncertainties
~\footnote{One measures the scalar
  dipole-polarisabilities in $10^{-4}\;\mathrm{fm}^3$, so that these units are
  dropped in the following.}, 
$\bar{\alpha}^p\approx11.0
$, $\bar{\beta}^p\approx2.8
$, with a theoretical uncertainty of $\approx1
$~\cite{Hi04}, see J.~McGovern's contribution to these proceedings for details.
A global analysis of the $28$ points for deuteron Compton
scattering gave
\begin{equation}
  \label{eq:neutronpols}
  \bar{\alpha}^s=11.3\pm0.7_\mathrm{stat}\pm0.6_\mathrm{Baldin}\pm1_\mathrm{th}
  \;,\; 
  \bar{\beta}^s =3.2\mp0.7_\mathrm{stat}\pm0.6_\mathrm{Baldin}\pm1_\mathrm{th}
\end{equation}
for the iso-scalar nucleon polarisabilities~\cite{Hi05b,Hi05} with the Baldin
sum rule $ \bar{\alpha}^{(s)}+\bar{\beta}^{(s)}=14.5\pm0.6$ as constraint.
Therefore, the proton and neutron polarisabilities are identical within
present theoretical and experimental uncertainties, as predicted by $\chi$EFT.
New data from MAXlab to improve the statistical uncertainties is being
analysed~\cite{myers}, and an experiment at HI$\gamma$S is approved.
Concurrently, a concerted effort is under way to reduce the theory-error using
higher orders in the chiral counting~\cite{allofus}.  Our goal is a
comprehensive approach to Compton scattering in the proton~\cite{Be02,Hi04},
deuteron~\cite{Be99,Be02,Hi05,Hi05b,Hi05a,Ch05} and ${}^3$He~\cite{Ch07} in
$\chi$EFT from zero energy to beyond the pion-production threshold.

Of particular interest are the $4$ so far practically un-determined
spin-polarisabilities $\gamma_{E1E1}$, $\gamma_{M1M1}$, $\gamma_{E1M2}$,
$\gamma_{M1E2}$ which parameterise the response of the nucleon-\emph{spin} to
the photon field, analogous to the Faraday effect of classical
Electrodynamics.
%
In these proceedings, we give a quick overview of our present investigations
to help extract spin-polarisabilities from polarised deuteron Compton
scattering.  As customary for proceedings, we apologise for our biased view
and refer to Refs.~\cite{Hi05b,Hi05,Ch05} and an upcoming
publication~\cite{hgds} for more detailed presentations and references.

\section{Ingredients and Observables}

\subsection{Dynamical Polarisabilities in $\chi$EFT}

Polarisabilities measure the global stiffness of the nucleon's internal
degrees of freedom against displacement in an electric or magnetic field of
definite multipolarity and non-vanishing frequency $\omega$ and are identified
\emph{at fixed energy} only by their different angular dependence. 
Nucleon Compton scattering provides thus a wealth of information about the
internal structure of the nucleon.
In contradistinction to most other electro-magnetic processes, it has however
only recently been analysed in terms of a multipole-expansion at fixed
energies~\cite{Hi04,He}.  Instead, one focused on the static polarisabilities,
i.e.~the values at zero photon energy. While quite different frameworks could
provide a consistent picture for the zero-energy values, the underlying
mechanisms are only properly revealed by their energy-dependence.
The complete set of dynamical polarisabilities does -- like all
multipole-decompositions -- not contain more or less information about the
nucleonic degrees of freedom than the Compton amplitudes. But the information
is more readily accessible and easier to interpret, as each mechanism leaves a
characteristic signature in a particular channel.

\begin{figure}[!htb]
\begin{center}
  \includegraphics*[width=0.95\linewidth]{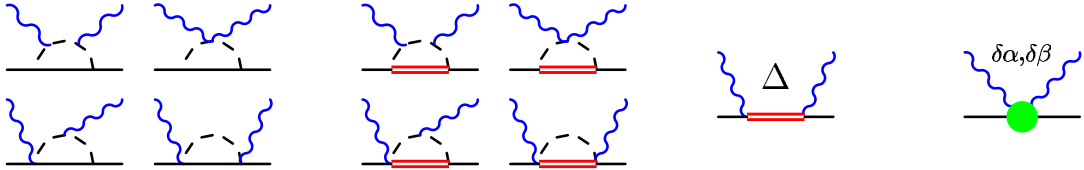}
  \caption{\label{fig:polas}The LO contributions to the nucleon
    polarisabilities. Left to right: pion cloud around the nucleon and
    $\Delta$; $\Delta$ excitations; short-distance effects. Permutations and
    crossed diagrams not shown.  From Ref.~\cite{Hi04}.}
\end{center}
\end{figure}

It is for example well-known that the $\Delta(1232)$ as the lowest nuclear
resonance leads by the strong $\gamma N\Delta$ $M1$-transition to a
para-magnetic contribution in the static magnetic dipole-polarisability
$\bar{\beta}_{\Delta} =+[7\dots13]$ and a characteristic resonance-shape as in
the Lorentz-Drude model of classical Electrodynamics. We therefore employ the
Chiral Effective Field Theory $\chi$EFT in which the $\Delta(1232)$ is
included as dynamical degree of freedom, in the ``Small Scale Expansion''
variant~\cite{He97}. The polarisability contributions at leading order (LO)
are listed in Fig.~\ref{fig:polas}. A $\pi^0$-pole contribution vanishes
because the deuteron is an iso-scalar.

At this order, the spin-polarisabilities are parameter-free predictions. But
as the observed static value $\bar{\beta}^p\approx 2$ is smaller by a factor
of $5$ than the $\Delta$ contribution, a strong dia-magnetic component must
exist. This fine-tuning at zero energy is unlikely to hold as $\omega$ is
varied: If dia- and para-magnetism are of different origin, they involve
different scales and hence different energy-dependences.  We sub-sume this
short-distance Physics which is at this order not generated by the pion or
$\Delta$ into two \emph{energy-independent} low-energy coefficients
$\delta\alpha,\;\delta\beta$. These ``off-sets'' are determined by data, and
the energy-dependence of the scalar polarisabilities is still a prediction of
$\chi$EFT.  Most notably even well below the pion-production threshold is the
strong energy-dependence induced into $\beta_{M1}(\omega)$ and
$\gamma_{M1M1}(\omega)$ by the $\Delta$-resonance. Its traditional
approximation as ``static-plus-small-slope'',
$\bar{\beta}+\omega^2\bar{\beta}_\nu$, is inadequate as low as $\omega\gtrsim
80$~MeV~\cite{Hi04}.  Not surprisingly, this contribution is most pronounced
at large momentum-transfers, i.e.~at backward angles. It resolves the ``SAL
puzzle'' of deuteron Compton scattering at
$94$~MeV~\cite{Hornidge,Hi05,Hi05a,Hi05b}, where widely varying iso-scalar
nucleon polarisabilities had been extracted, in disagreement with data taken
at lower energies.

\subsection{Embedding the Nucleon in the Deuteron}

\begin{figure}[!htbp]
  \begin{center}
    \includegraphics*[width=0.5\linewidth]{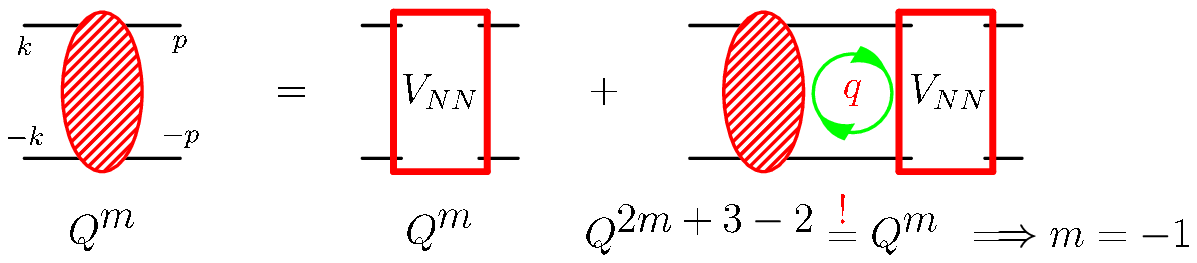}
  \caption{\label{fig:consistency}
    On the consistency of $NN$ power-counting in $\chi$EFT. From
    Ref.~\cite{hg}.}
\end{center}
\end{figure}

Neutrons properties are usually extracted from data taken on few-nucleon
systems by dis-entangling nuclear-binding effects. $\chi$EFT allows to
subtract two-body contributions of meson-exchange currents and of
wave-function dependence from data with minimal theoretical prejudice and with
an estimate of the theoretical uncertainties. A consistent description must
also give the correct Thomson limit, an exact low-energy theorem which in turn
follows from gauge invariance~\cite{Friar}. Its verification is
straight-forward in the 1-nucleon sector, where the amplitude is perturbative.
But the two-nucleon amplitude must be non-perturbative to accommodate the
shallow bound-state: All terms in the LO Lippmann-Schwinger equation of
$NN$-scattering, Fig.~\ref{fig:consistency}, including the potential, must be
of the same order when all nucleons are close
to their non-relativistic mass-shell. Otherwise, one of them could be treated
as perturbation of the others and a low-lying bound-state would be absent.
Picking the nucleon-pole in the energy-integration $E\sim\frac{Q^2}{2M}$ leads
therefore to the consistency condition that the $NN$-scattering amplitude
$T_{NN}$ must be of order $Q^{-1}$, irrespective of the potential used.  Here,
$Q$ is a typical low-momentum scale of the process under consideration,
e.g.~the inverse $\mathrm{S}$-wave scattering length.  The relative strength
of forces and potentials in $\chi$EFT is therefore not just determined by
counting the number of momenta. This has long been recognised in ``pion-less''
EFT, but is only an emerging communal wisdom in the chiral
version~\cite{hg,NoggaBirse,hgpc}, see also Birse's contribution to these
proceedings.

In deuteron Compton scattering, this mandates to include $T_{NN}$ whenever
both nucleons propagate close to their mass-shell between photon absorption
and emission, i.e.~when the photon energy $\omega\lesssim 50\;\mathrm{MeV}$
does not suffice to knock a nucleon far off its mass-shell~\cite{hg}.
\begin{figure}[!htbp]
  \begin{center}
    \includegraphics*[width=\linewidth]{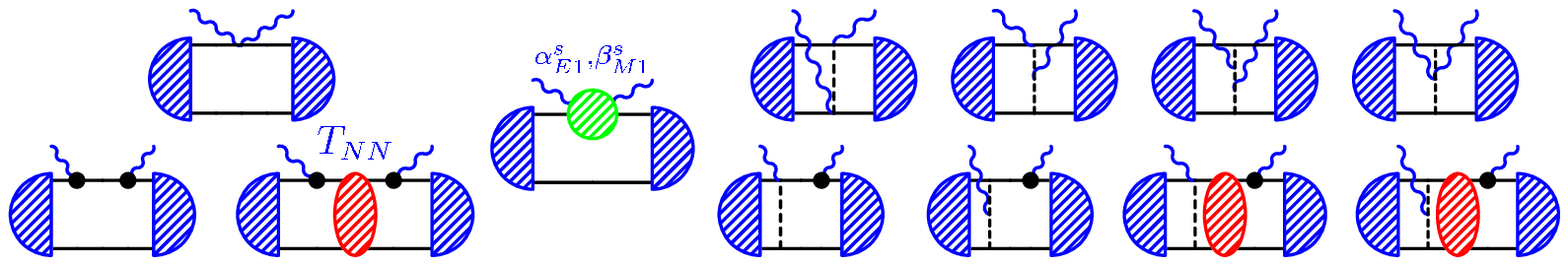}
    \caption{\label{fig:dgraphs} Deuteron Compton scattering in $\chi$EFT to
      NLO. Left: one-body part (dot: electric/magnetic coupling; blob: nucleon
      polarisabilities of Fig.~\protect\ref{fig:polas}). Right: two-body part
      (pion-exchange currents).  Permutations and crossed graphs not shown.
      From Ref.~\protect\cite{Hi05b}.  }
\end{center}
\end{figure}
Figure~\ref{fig:dgraphs} lists the contributions to Compton scattering off the
deuteron to next-to-leading order NLO in $\chi$EFT. At higher photon energies
$\omega\gtrsim 60\;\mathrm{MeV}$, the nucleon is kicked far enough off its
mass-shell, $E\sim Q$, for the amplitude to become perturbative. This is
intuitively clear, as the struck nucleon has only a very short time
($\sim1/\omega$) to scatter with its partner before the second photon has to
be radiated to restore the coherent final state. The diagrams which contain
$T_{NN}$ in Fig.~\ref{fig:dgraphs} are therefore less important for larger
$\omega$, together with some of the other diagrams. Indeed, the nucleon
propagator becomes static and scales as $1/Q\sim1/\omega$, with each
re-scattering process in $T_{NN}$ suppressed by an additional power of $Q$.
However, $NN$-rescattering practically eliminates even at these high energies
the dependence on the potential used to produce the deuteron wave-function and
$NN$-rescattering matrix. We implemented rescattering by the Green's function
method described in~\cite{Karakowski,Lvov,Hi05,Hi05b}. The calculation is
parameter-free after fitting the scalar polarisabilities with the result
quoted in eq.~(\ref{eq:neutronpols})~\cite{Hi05,Hi05b}.

\subsection{Deuteron Observables}

\begin{figure}[!htb]
\begin{center}
\small{{linpol.~$\gamma$, unpol.~d:}}
{$\displaystyle\left.\frac{\dd\sigma}{\dd\Omega}\right|^\mathrm{lin}_x$} \hq
\includegraphics*[width=0.1\linewidth]{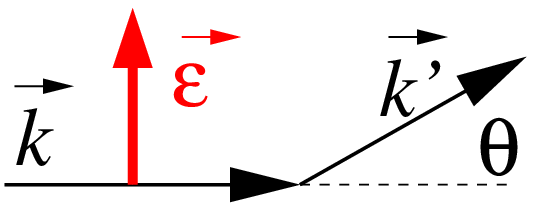}
\hq\hq,\hq\hq
{$\displaystyle\left.\frac{\dd\sigma}{\dd\Omega}\right|^\mathrm{lin}_y$} \hq
\includegraphics*[width=0.1\linewidth]{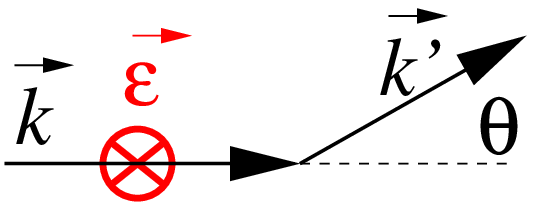}\\[2ex]
\parbox{10ex}{{circpol.~$\gamma$,\\ vecpol.~d}}:
{$\displaystyle\Delta_x^\mathrm{circ}
$}
\hq\parbox{0.35\linewidth}{
\includegraphics*[width=\linewidth]{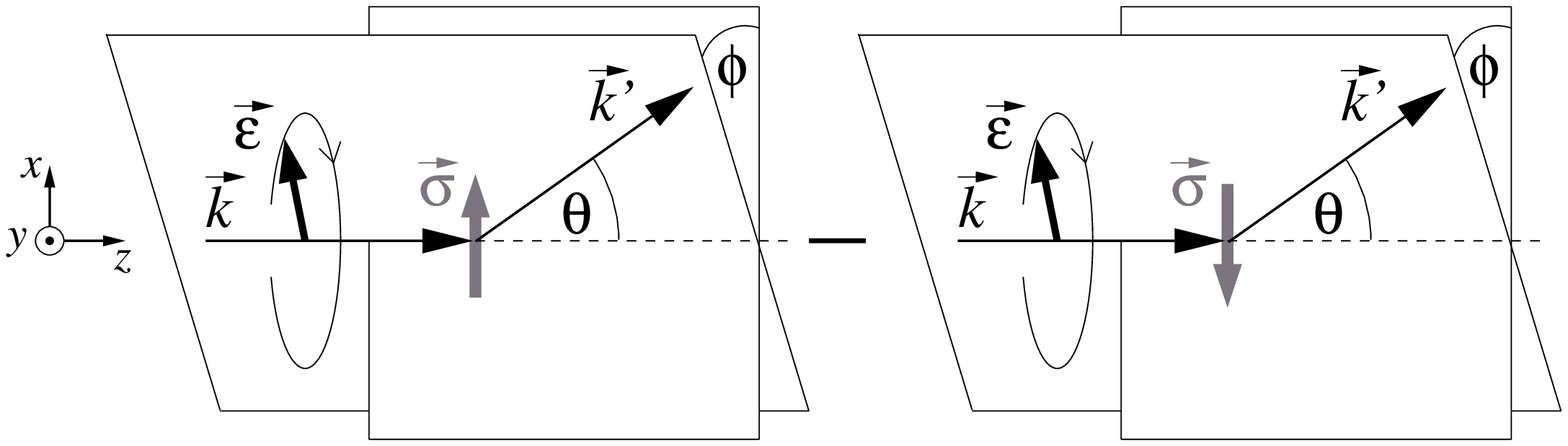}}
\hq\hq,\hq\hq
{$\displaystyle\Delta_z^\mathrm{circ}
$}
\hq\parbox{0.35\linewidth}{
\includegraphics*[width=\linewidth]{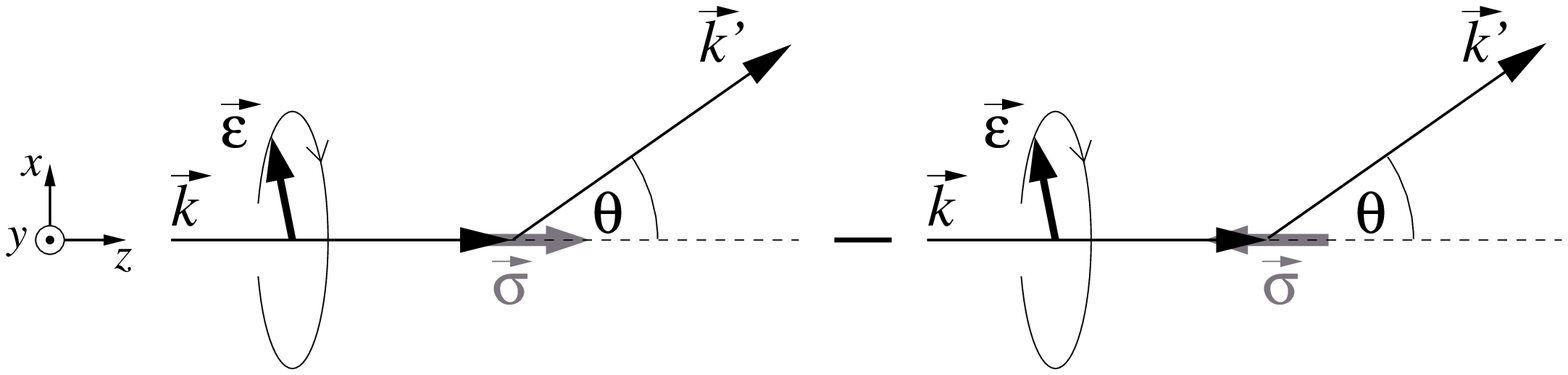}}\\[2ex]
\parbox{10ex}{{linpol.~$\gamma$,\\ vecpol.~d}}:
{$\displaystyle\Delta_x^\mathrm{lin}
$}
\hq\parbox{0.35\linewidth}{
\includegraphics*[width=\linewidth]{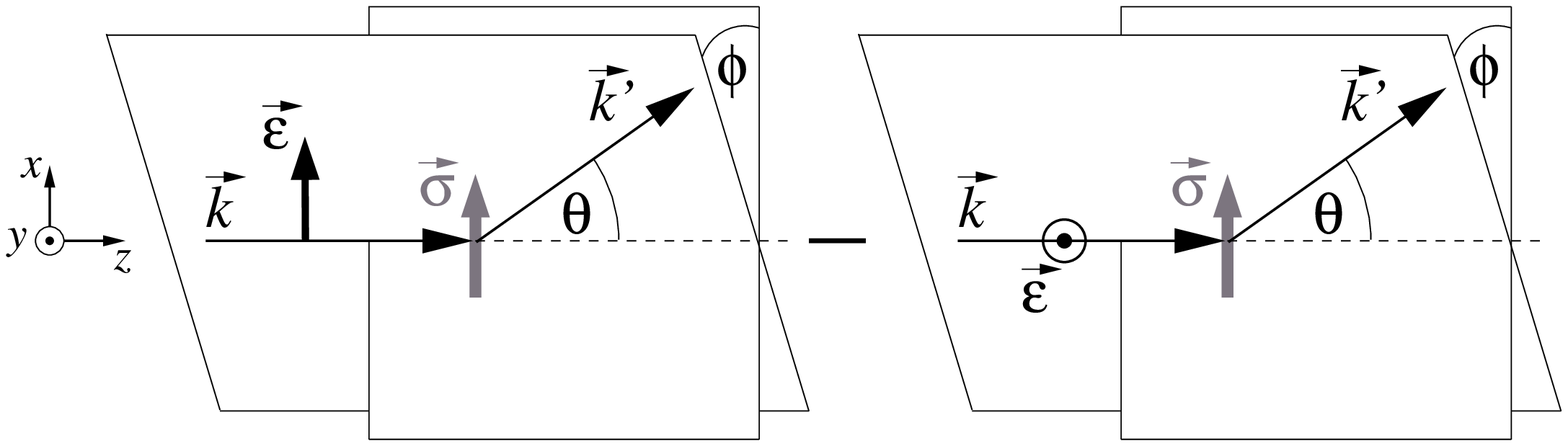}}
\hq\hq,\hq\hq
{$\displaystyle\Delta_z^\mathrm{lin}
$}
\hq\parbox{0.35\linewidth}{
\includegraphics*[width=\linewidth]{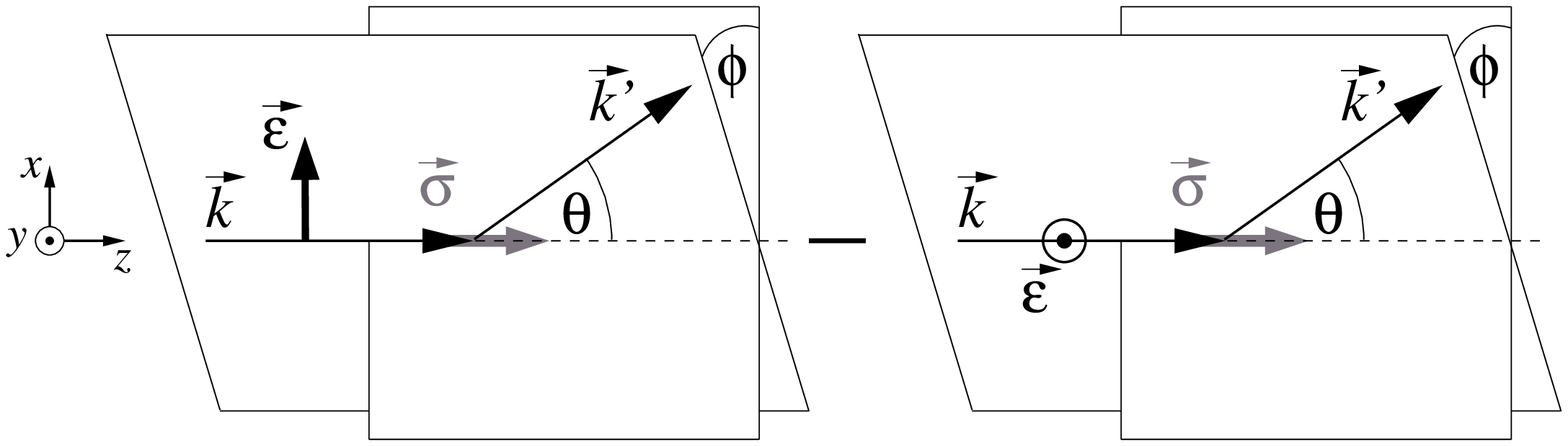}}
\caption{\label{fig:observables}Definition of observables for singly and double
  polarised cross-sections. }
\end{center}
\end{figure}

Besides the unpolarised cross-section of Refs.~\cite{Hi05,Hi05a,Hi05b}, new
techniques allow measuring observables with polarised beams and/or targets.
For a linearly polarised beam and unpolarised deuteron,
$\left.\frac{\dd\sigma}{\dd\Omega}\right|_x^\mathrm{lin}$ is the differential
cross-section for photon polarisation in the scattering plane, and
$\left.\frac{\dd\sigma}{\dd\Omega}\right|_y^\mathrm{lin}$ for perpendicular
polarisation. The double polarised observables $\Delta$ involve a
vector-polarised deuteron and a circularly or linearly polarised photon.
Often, experiments discuss asymmetries $\Sigma$, i.e.~cross-section
differences divided by their sums, to cancel systematic effects.
Figure~\ref{fig:observables} gives a pictorial representation of the
observables considered, in lieu of lengthy formulae.

Cross-section differences and asymmetries for $6$ observables, depending on
$6$ dipole polarisabilities and $3$ kinematic variables (photon energy
$\omega$ and scattering angles $\theta$ and $\phi$) in the cm and lab frame,
and additional constraints like the Baldin sum rule and the forward and
backward spin-polarisabilities, provide a cornucopia of information which
cannot adequately be conveyed in a short article. We therefore focus only on
some prominent examples here and note that in order to aide in planning new
experiments, the results for all observables are available as an interactive
\emph{Mathematica 7.0} notebook from Grie{\ss}hammer (hgrie@gwu.edu). It
produces both tables and plots of energy- and angle-dependences from $10$ to
$\approx120$~MeV of all the asymmetries and cross-section differences, as well
as of the total cross-section, in both the cm and lab systems, including their
sensitivities to varying the spin-independent and spin-dependent
polarisabilities independently.

\section{Significance of  the $\Delta(1232)$ and $NN$ Rescattering on Polarisation Observables}
\label{sec:comparison}

\begin{figure}[!htb]
\begin{center}
\includegraphics*[width=0.31\linewidth]{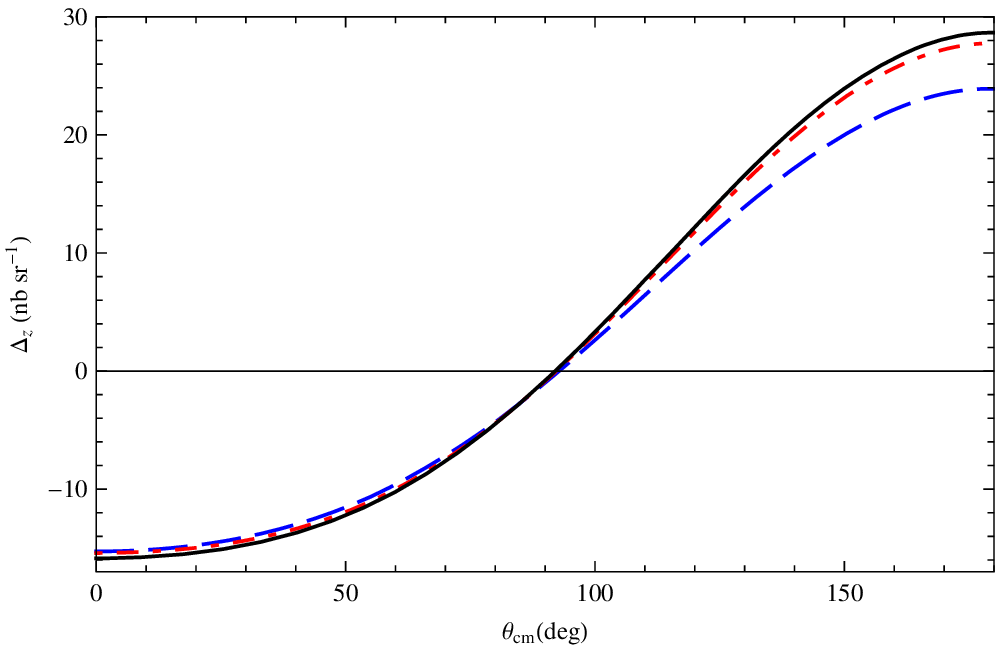}
\includegraphics*[width=0.31\linewidth]{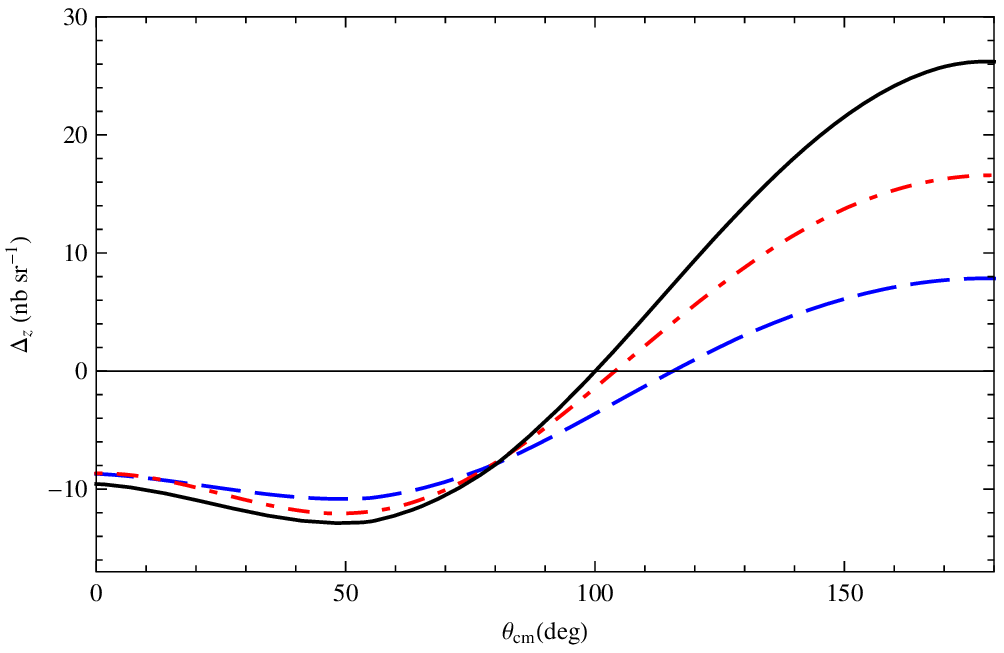}\\
\includegraphics*[width=0.31\linewidth]{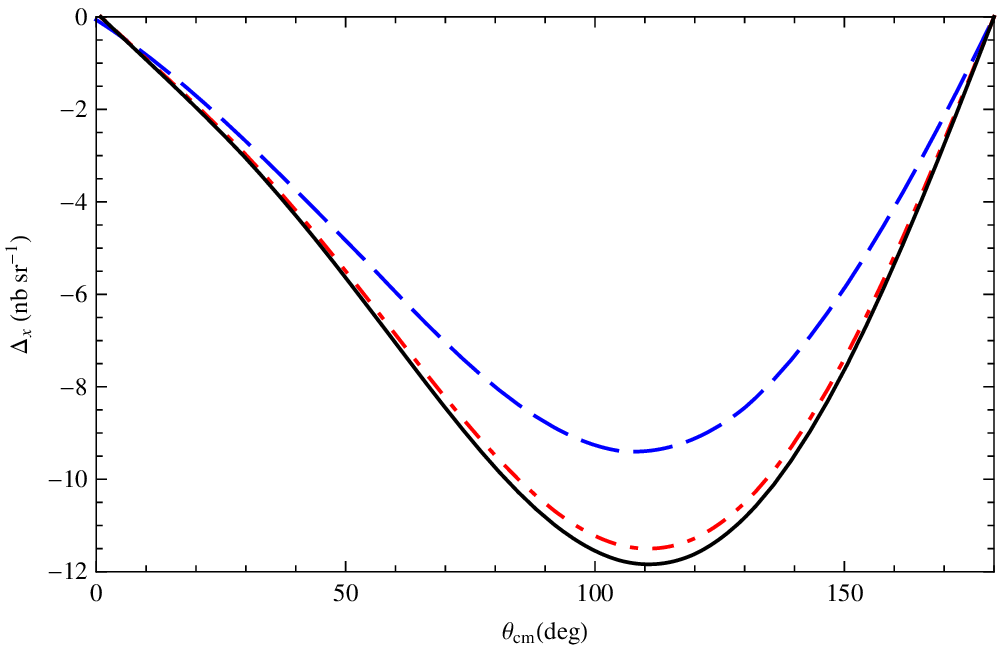}
\includegraphics*[width=0.31\linewidth]{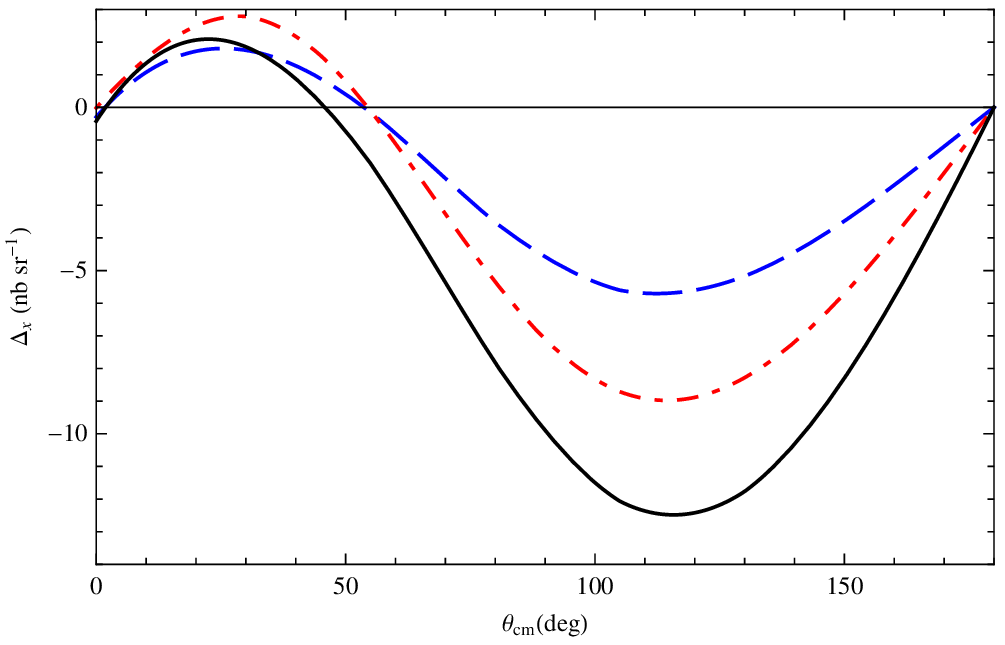}
\caption{\label{fig:comp}Effects of $\De(1232)$ and of resumming $NN$
  intermediate states on the double polarisation observables $\De_z$ (top) and
  $\De_x$ (bottom). Left: $\w_\mathrm{lab}=45$ MeV; right:
  $\w_\mathrm{lab}=125$ MeV. 
  Dashed (blue): ${\mathcal O} (Q^3)$ HB\cpt calculation without dynamical
  $\Delta(1232)$ or rescattering. Solid (black): ${\mathcal O}
  (\varepsilon^3)$ calculation with both intermediate $NN$ rescattering and
  dynamical $\Delta(1232)$. Dot-dashed (red) on left: only $NN$-rescattering
  added, no dynamical $\Delta(1232)$. Dot-dashed (red) on right: only
  dynamical $\Delta(1232)$ added, no $NN$-rescattering.  }
\end{center}
\end{figure}

We first analyse the $\De(1232)$ and intermediate $NN$ rescattering
contributions on polarisation observables.  Figure~\ref{fig:comp} compares
double-polarisation observables within different schemes. The upper (lower)
row shows the parallel (perpendicular) polarisation asymmetry
$\Delta_z^\mathrm{circ}$ ($\Delta_x^\mathrm{circ}$) with circularly polarised
photons. The left (right) panels are for $\w_\mathrm{lab}=45$~MeV ($125$~MeV).
As for unpolarised observables~\cite{Hi05,Hi05a}, the $\De(1232)$ does not
contribute appreciably at $45$~MeV, but the observable is still ruled by
including intermediate $NN$ rescattering for the correct Thomson limit. In
contradistinction, the $\De$ and intermediate $NN$-rescattering are equally
significant at $125$~MeV. 
%

Like in Refs.~\cite{Hi05,Hi05a} for unpolarised observables, we find that both
the $\De(1232)$ and intermediate $NN$ rescattering are necessary ingredients
to identify polarisation observables for reliably extracting polarisabilities
from zero to $125$~MeV. We also checked that dependence on the potential used
to produce the deuteron wave-function and $NN$-rescattering matrix is
irrelevant.

\section{Results}
\label{sec:results}

\begin{figure}[!htb]
\begin{center}
\includegraphics*[width=.31\linewidth]{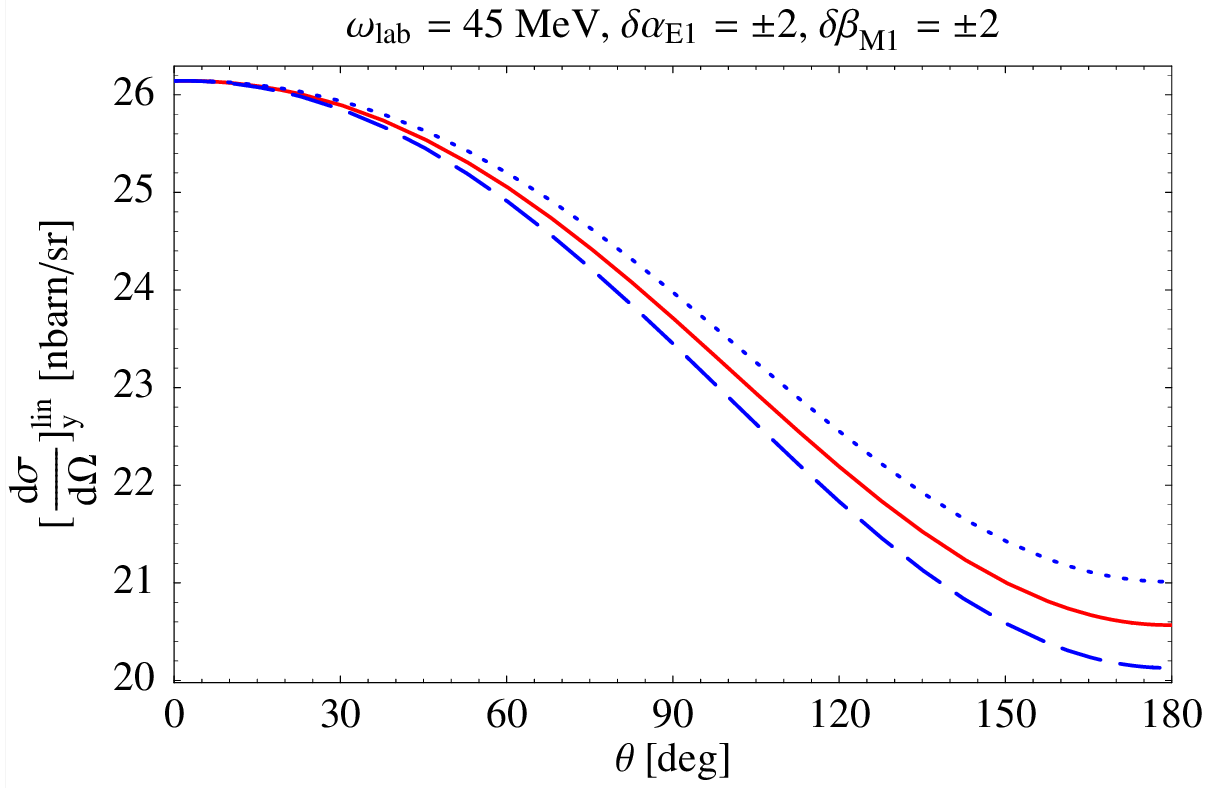}
\hfill
\includegraphics*[width=.31\linewidth]{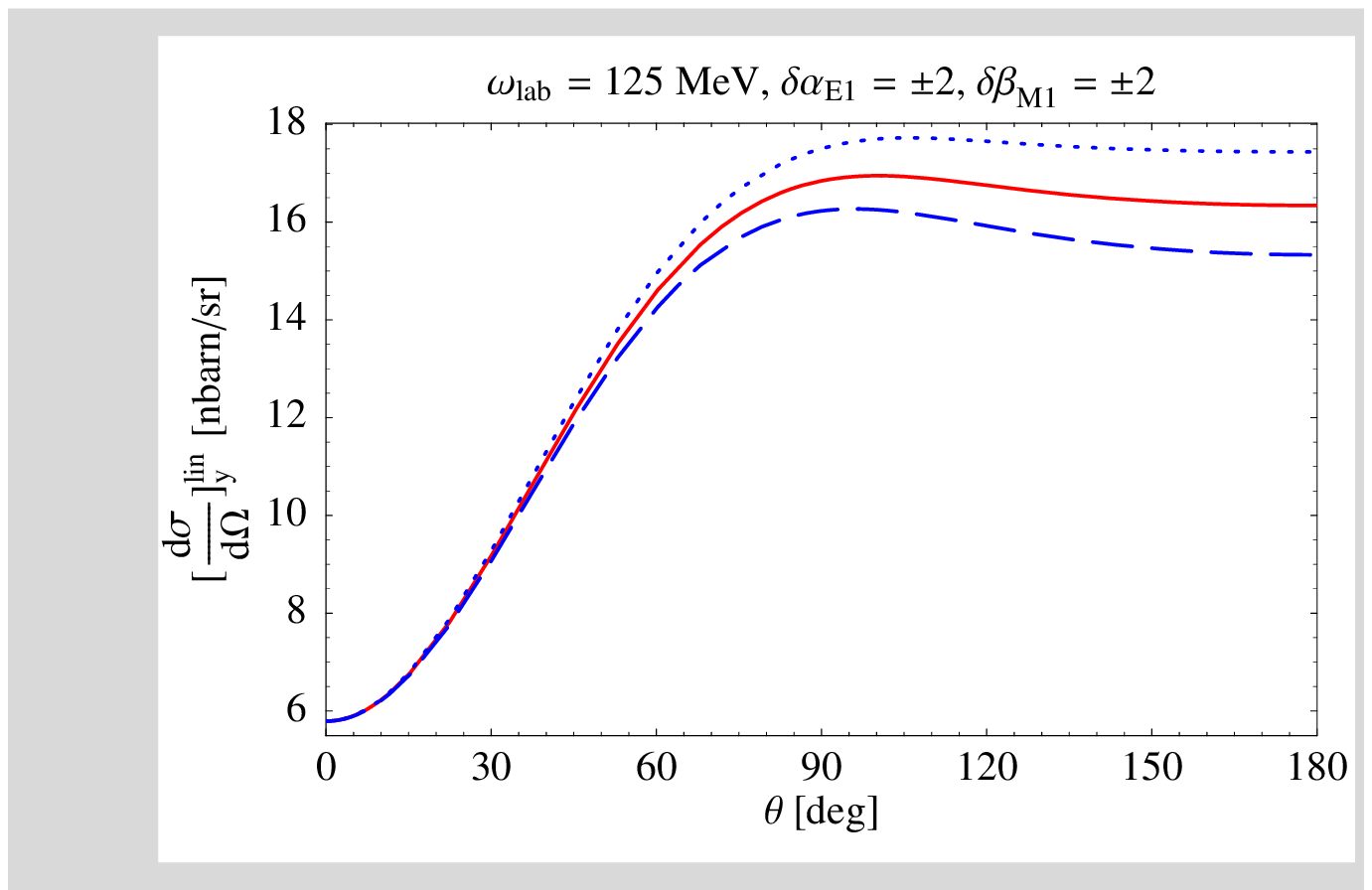}
\hfill
\includegraphics*[width=.31\linewidth]{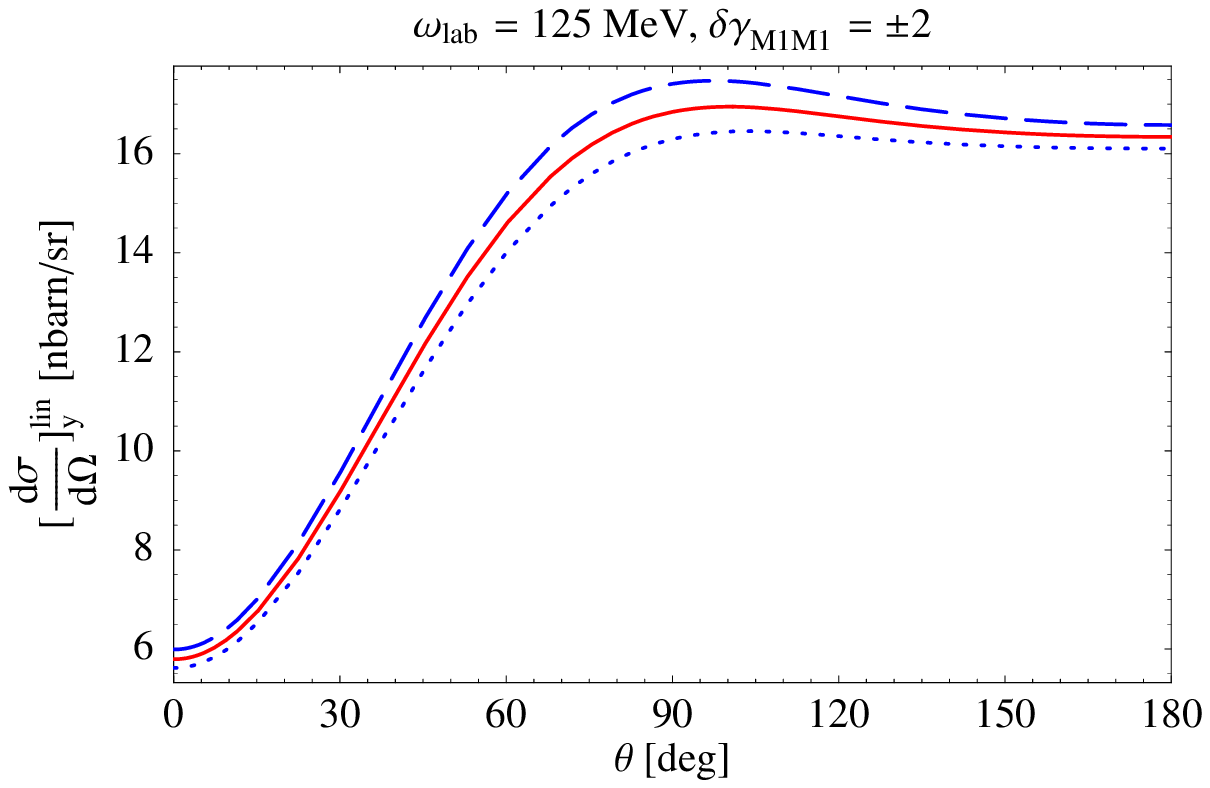}
\caption{\label{fig:dcsy} Differential cross-sections with photons
  linearly-polarised along the $y$-axis. Left and centre: Scalar
  polarisabilities $\bar{\alpha}^{(s)}$ and $\bar{\beta}^{(s)}$ are each
  varied by $\pm2$ units, while the sum is constrained by the iso-scalar
  Baldin sum rule. Right: $\ga_{M1M1}$ varied by $\pm2$ units at
  $\w_\mathrm{lab}=125$ MeV.}
\end{center}
\end{figure}

\begin{figure}[!htb]
\vspace{-2ex}

\begin{center}
\includegraphics*[width=.31\linewidth]{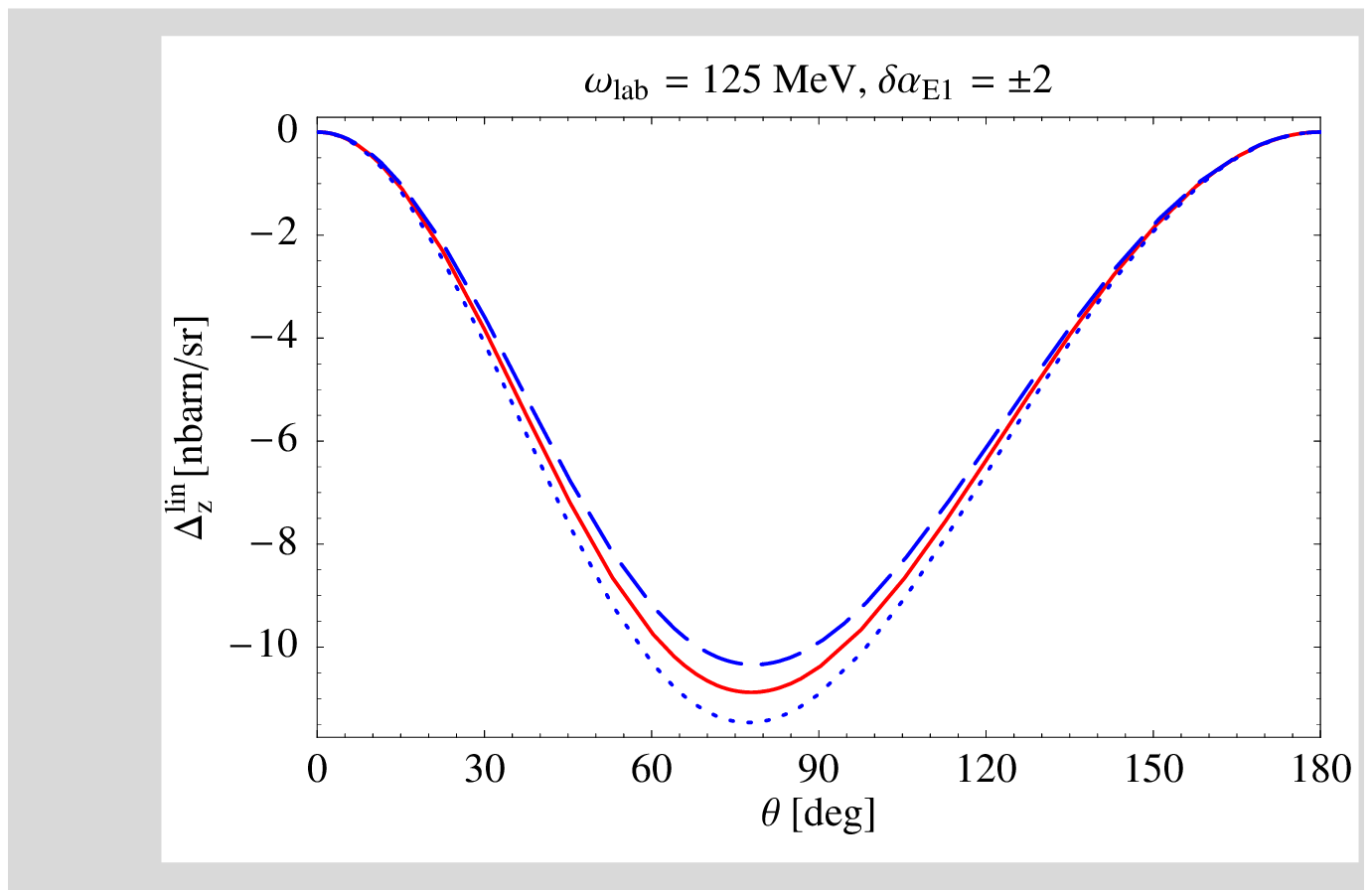}
\hfill
\includegraphics*[width=.31\linewidth]{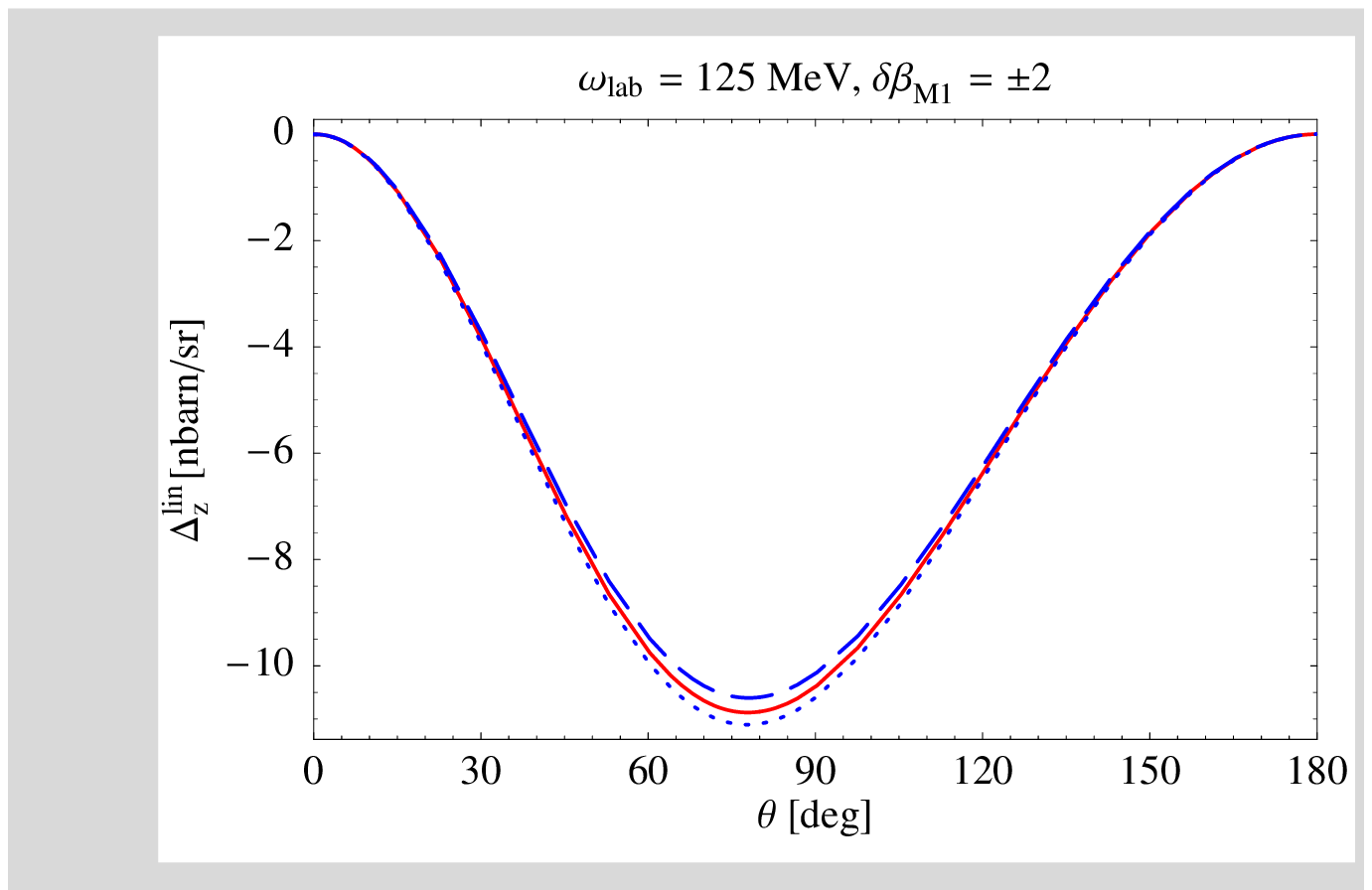}
\hfill
\includegraphics*[width=.31\linewidth]{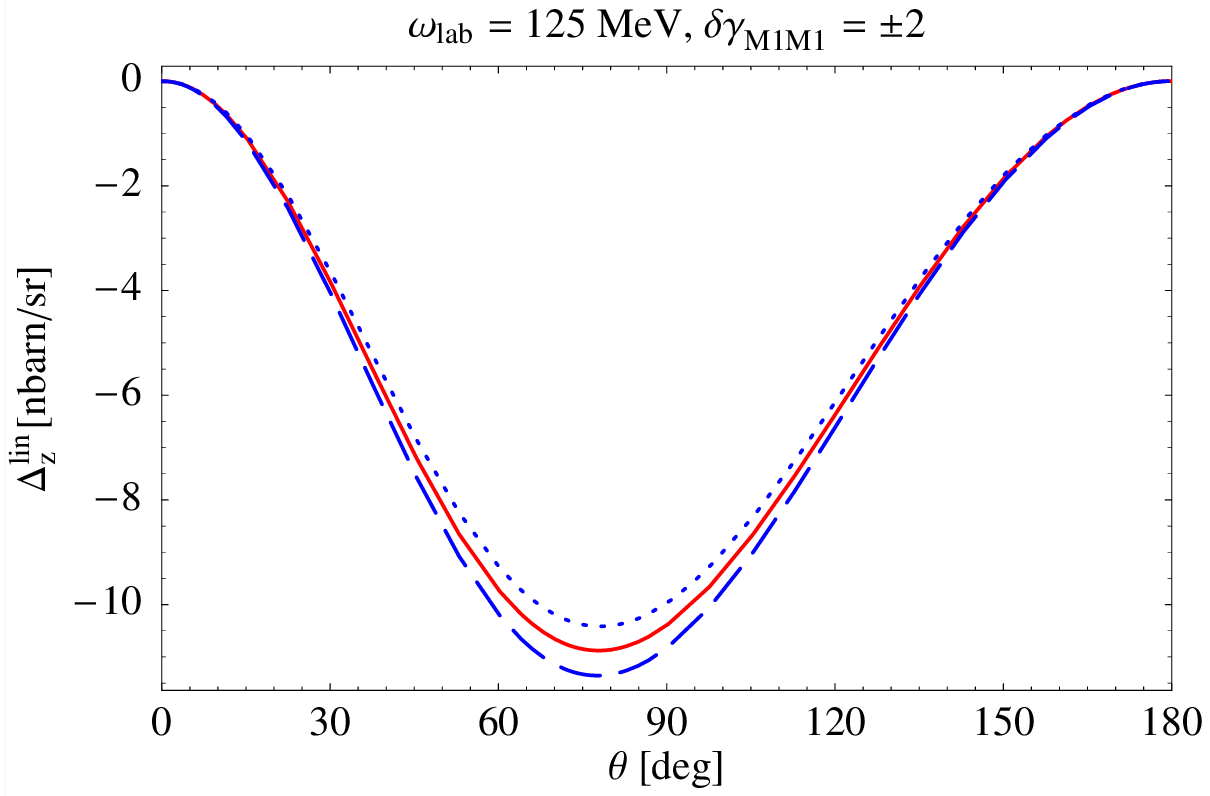}
\caption{\label{fig:deltazlin} The double-polarisation asymmetry
  $\Delta_z^\mathrm{lin}$ with linearly-polarised photons at $\w_\mathrm{lab}=125$
  MeV. Left/centre: $\bar{\alpha}^{(s)}$ and $\bar{\beta}^{(s)}$
  varied by $\pm2$ units, no Baldin constraint.
  Right: variation of $\ga_{M1M1}$ by $\pm2$.}
\end{center}
\end{figure}

We now identify selected polarisation observables which are helpful in
extracting in particular spin-polarisabilities. For an unpolarised target,
$\left.\frac{\dd\sigma}{\dd\W}\right|_y^{lin}$ is at $45$~MeV (lab)
appreciable sensitive only to $\alpha_{E1}$ and $\beta_{M1}$, see
Fig.~\ref{fig:dcsy}.  At 125 MeV (lab) however, the sensitivity to
$\ga_{M1M1}$ is large and comparable to that of $\al_{E1}-\be_{M1}$.
Amongst double-polarisation observables at $\w_\mathrm{lab}=$ 125~MeV, there
is also appreciable sensitivity to the polarisabilities in the linear photon
polarisation asymmetry $\De_z^{lin}$, see Fig.~\ref{fig:deltazlin}. The
dependences on $\al_{E1}$, $\be_{M1}$ and $\ga_{M1M1}$ are again comparable.
Finally, the circular photon polarisation asymmetry $\De_x^{circ}$ at
$\w_\mathrm{lab}=$ 125~MeV, shows large and comparable sensitivity on
$\al_{E1}$, $\be_{M1}$ and now $\ga_{E1E1}$, but only minor sensitivity on the
other spin-polarisabilities.

\begin{figure}[!htb]
\vspace{-2ex}
\begin{center}
\includegraphics*[width=0.31\linewidth]{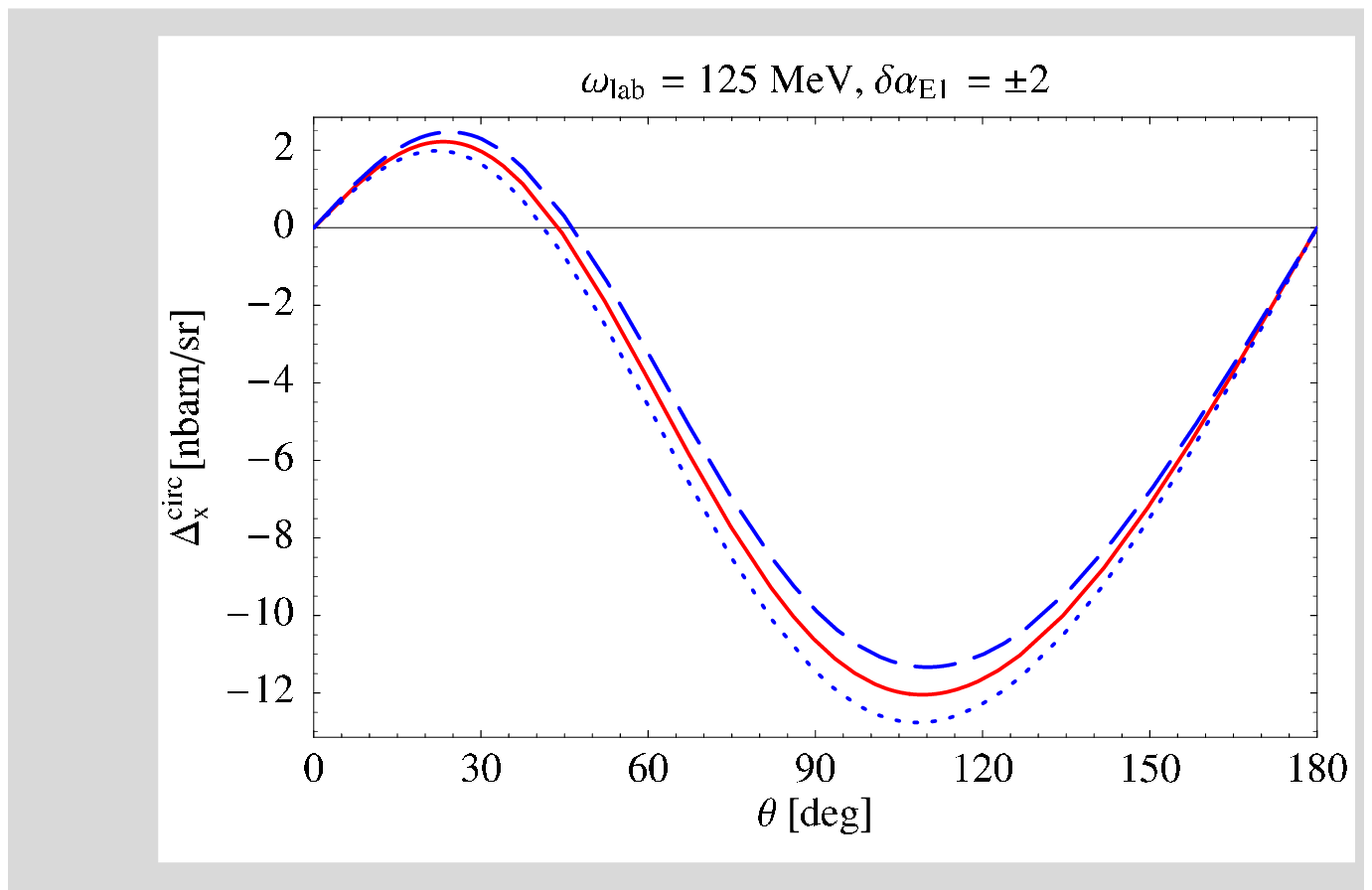}
\hq\hq
\includegraphics*[width=0.31\linewidth]{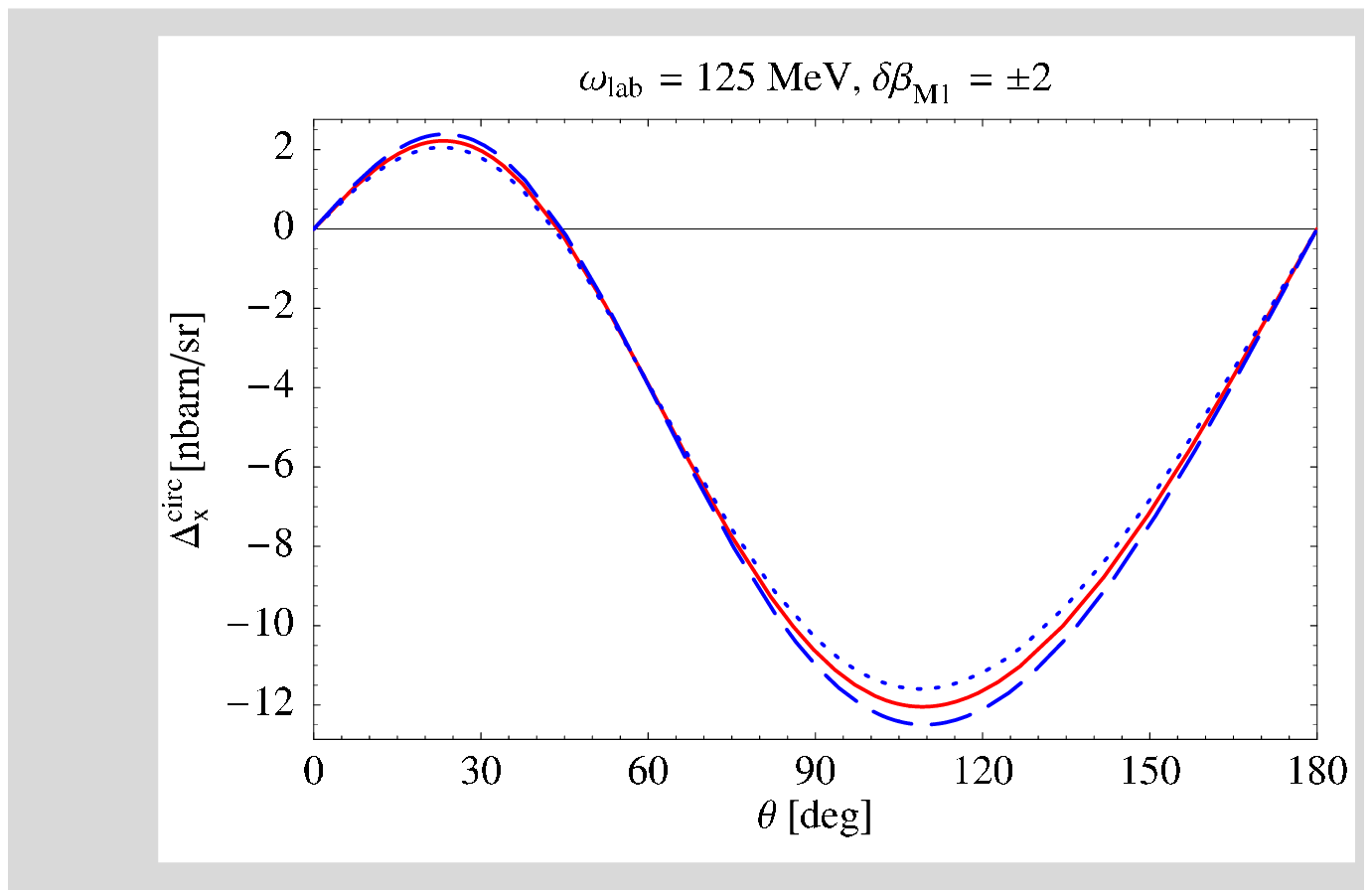}
\hq\hq
\includegraphics*[width=0.31\linewidth]{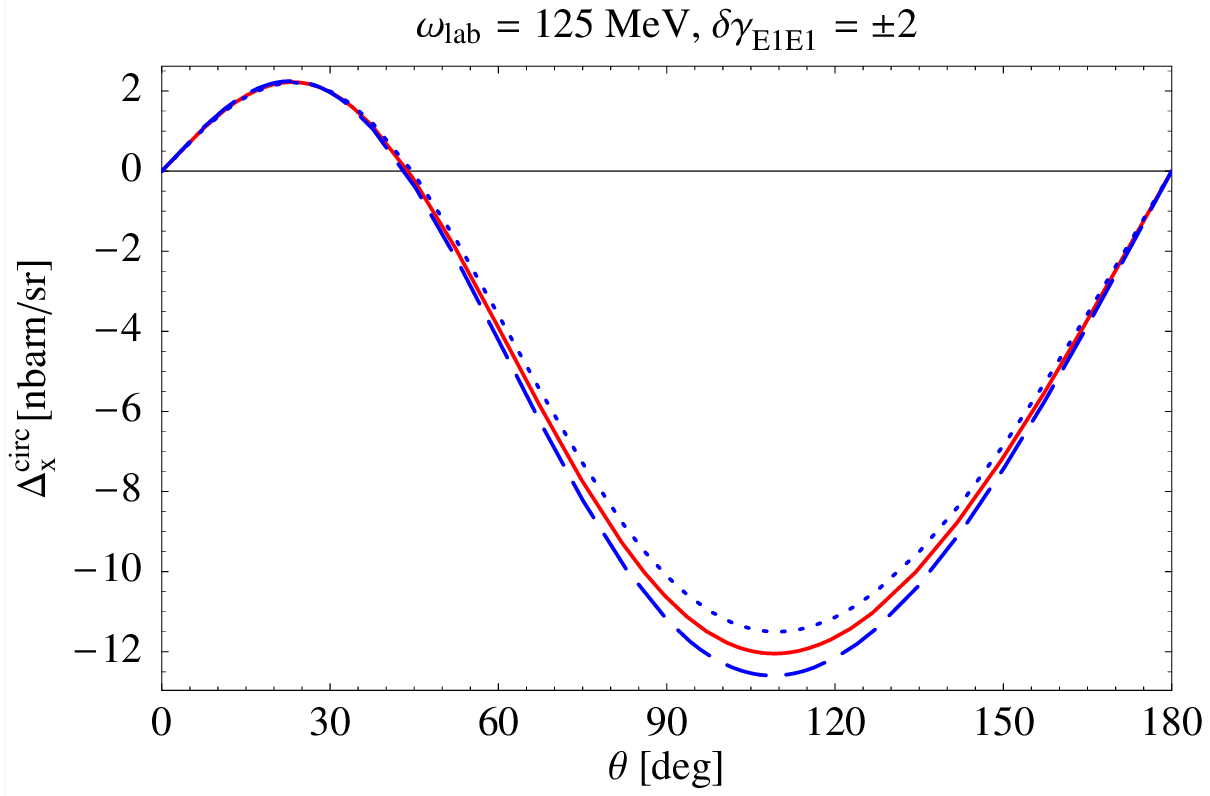}\\[1ex]
\includegraphics*[width=0.31\linewidth]{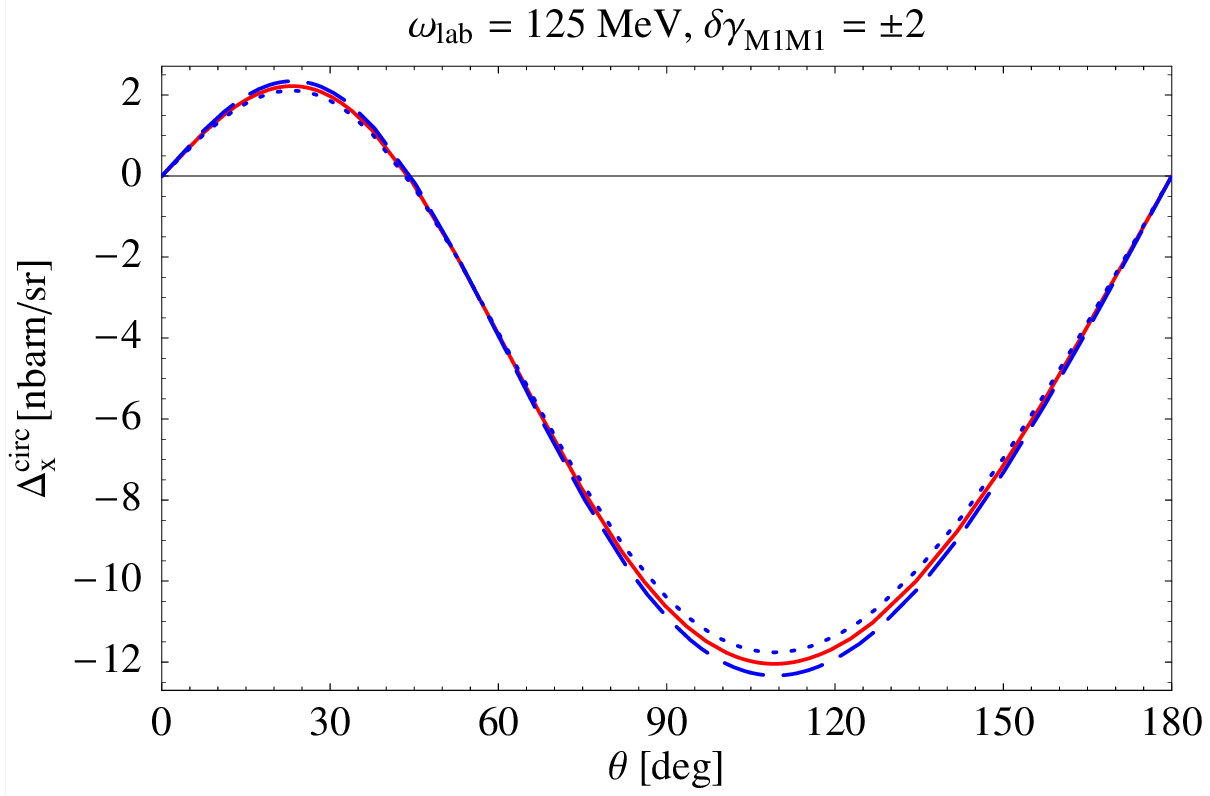}
\hq\hq
\includegraphics*[width=0.31\linewidth]{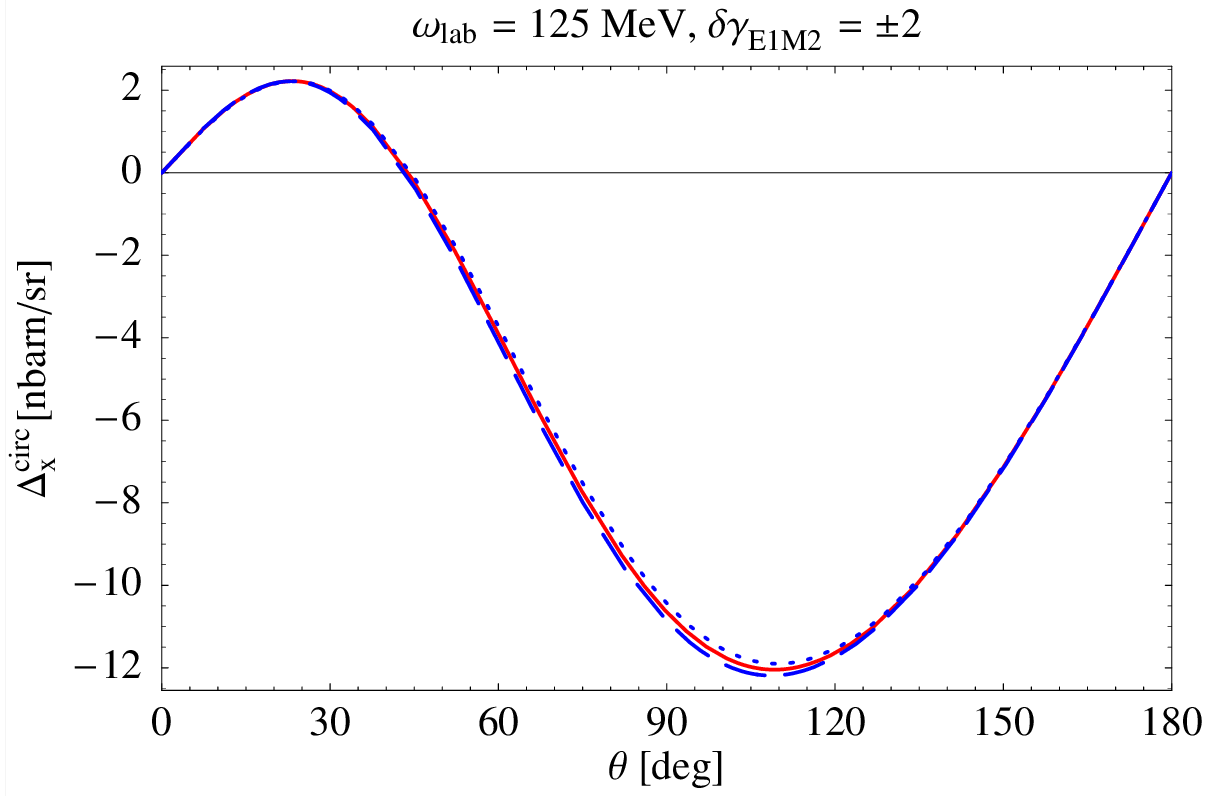}
\hq\hq
\includegraphics*[width=0.31\linewidth]{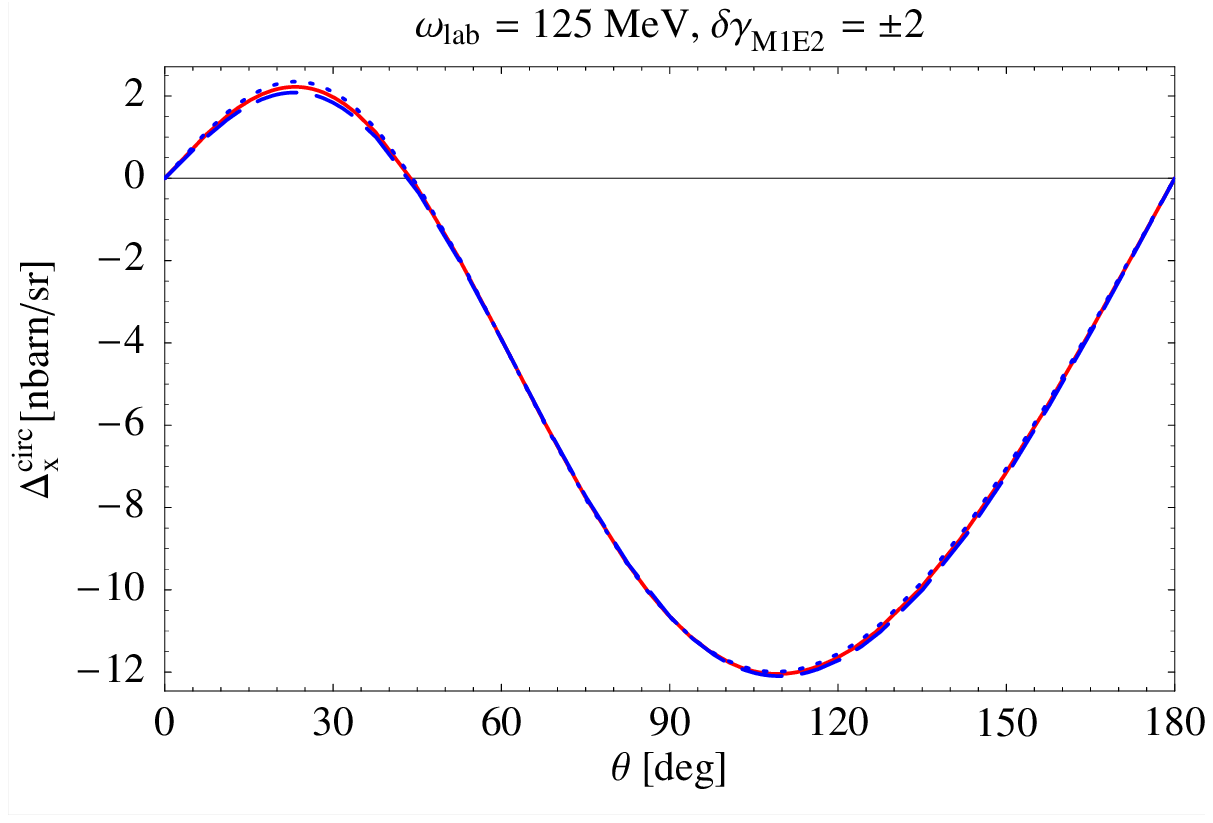}
\caption{\label{fig:deltax} The double-polarisation asymmetry
  $\Delta_z^\mathrm{circ}$ with circularly-polarised photons at $\w_\mathrm{lab}=125$
  MeV. From top left to bottom right, variation by $\pm2$ units of
  $\alpha_{E1}$, $\beta_{M1}$, $\gamma_{E1E1}$, $\gamma_{M1M1}$,
  $\gamma_{E1M2}$, $\gamma_{M1E2}$.  }
\end{center}
\end{figure}

\begin{figure}[!htb]
\vspace{-2ex}
\begin{center}
\parbox{0.13\linewidth}{
\includegraphics*[width=\linewidth]{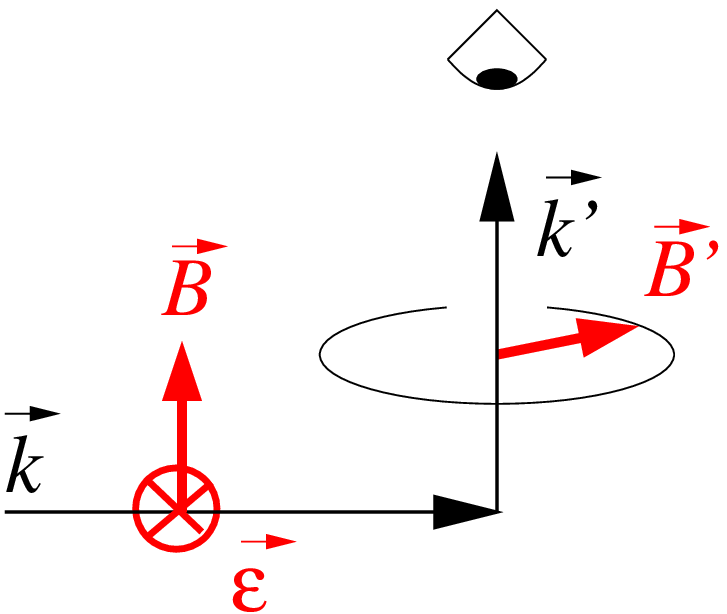}}
\hq\hq\hq\hq
\parbox{0.64\linewidth}{
\includegraphics*[width=0.48\linewidth]{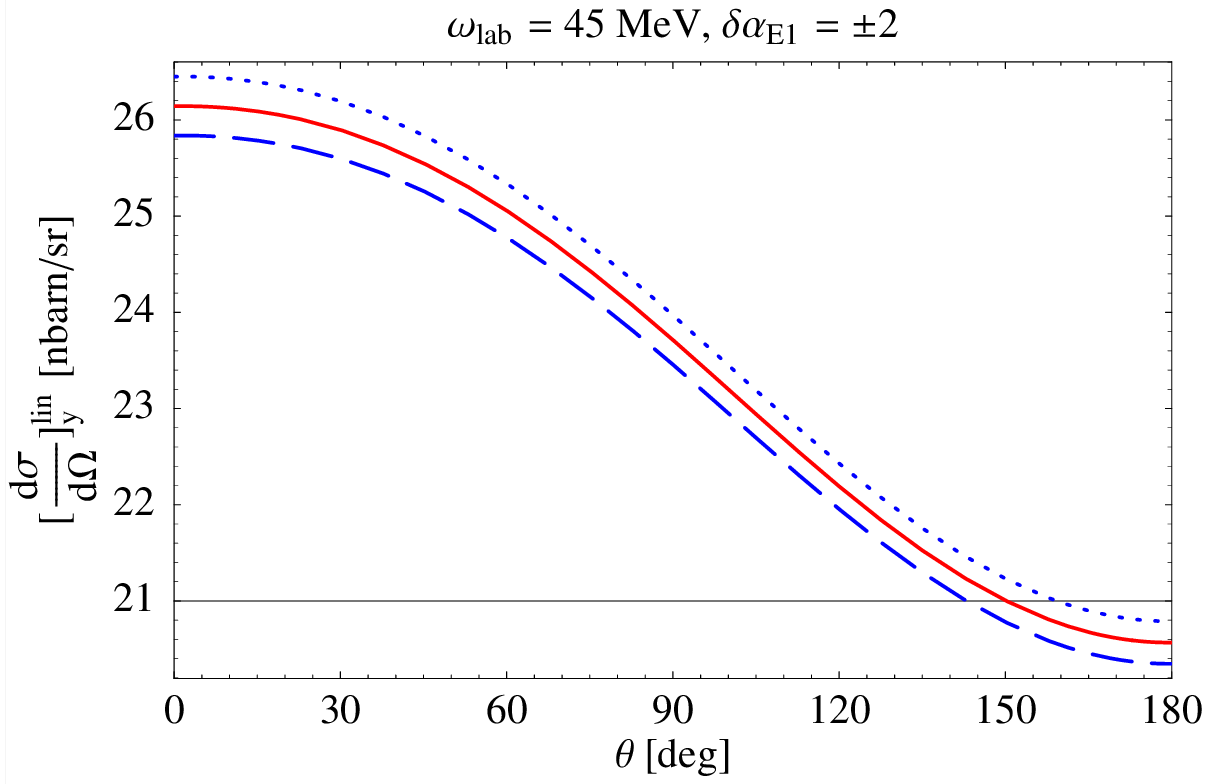}
\hq
\includegraphics*[width=0.48\linewidth]{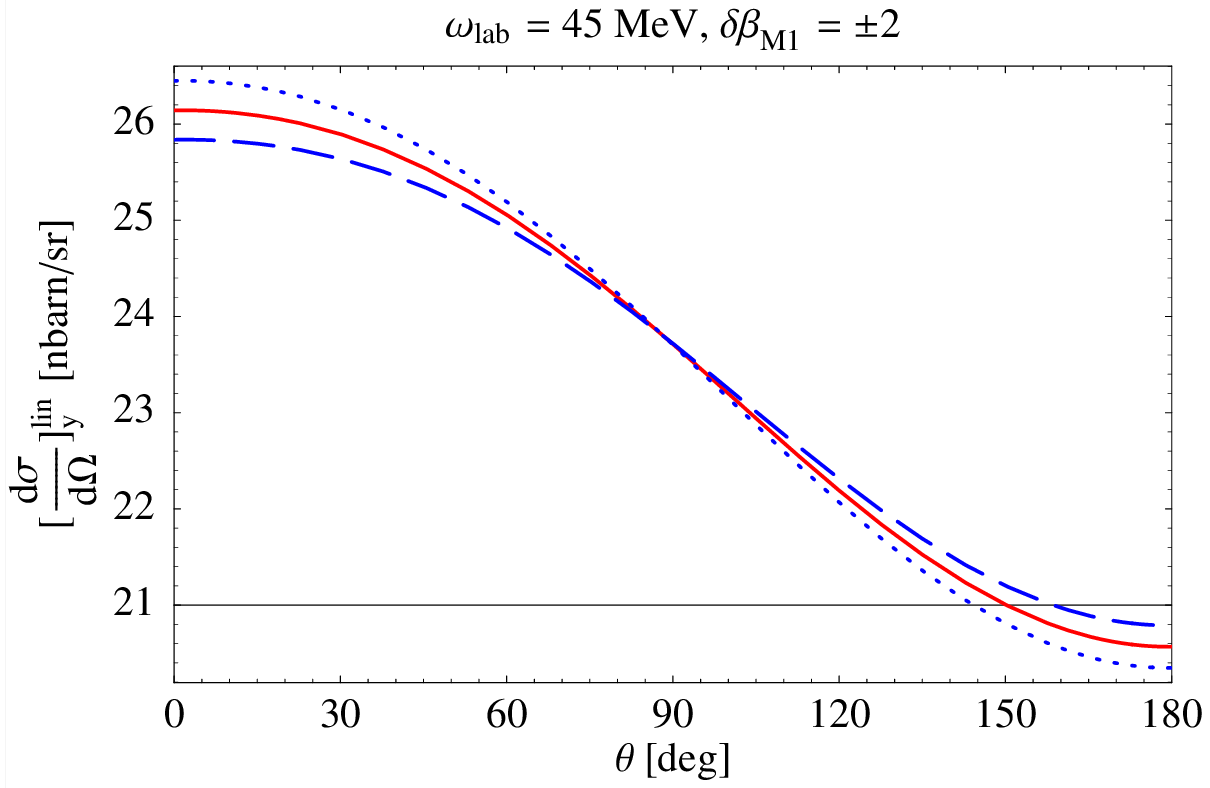}}
\caption{\label{fig:switchingoffpols} Left: Configuration under which a
  induced dipole cannot radiate an $M1$ photon into the detector.
  Centre/right: $\left.\frac{\dd\sigma}{\dd\Omega}\right|_y^\mathrm{lin}$ when
  $\alpha_{E1}$/$\beta_{M1}$ is varied by $\pm2$. Notice the
  $\beta$-independence at $90^\circ$.}
\end{center}
\end{figure}

Thus, it is imperative that the values of the electric and magnetic
polarisabilities be better extracted so as not to taint any extraction of the
spin polarisabilities. While no observable is sensitive only to one dipole
polarisability, closely inspecting (\ref{polsfromints}) reveals configurations
in which one (or more) polarisabilities do \emph{not} contribute for nucleon
Compton scattering. In the example of Fig.~\ref{fig:switchingoffpols}, a
photon is scattered on an unpolarised nucleon in the cm frame such that the
linear photon polarisation is perpendicular to the scattering plane,
cf.~\cite{Maximon:1989zz}. A detector under $90^\circ$ can therefore not
detect $M1$ photons radiated from the induced magnetic dipole in the nucleon.
As demonstrated in the figure, this hold even when the relative motion of the
nucleon in the deuteron is taken into account~\cite{hgds}. Similar
configurations can be identified for other dipole polarisabilities.

\section{Concluding Questions}

Our results indicate that some observables of single- and double-polarised
deuteron Compton scattering can be used to directly extract some of the so-far
nearly un-determined spin-polarisabilities.  However, as these are
higher-order relative to the electric and magnetic polarisabilities, the
former can only be extracted reliably when the scalar polarisabilities are
known with better accuracy. With this goal, an experiment at MAXlab is being
analysed as we speak, and an experiment at HI$\gamma$S is approved.
Concurrently, we are improving the theoretical accuracy by including higher
orders in $\chi$EFT~\cite{allofus}. To find all nucleon spin-polarisabilities,
one therefore needs to work through:
\begin{itemize}
\item[(1)] A set of relatively low-energy experiments,
  $\omega\lesssim80$~MeV, where the spin-polarisabilities are
  negligible but the scalar polarisabilities can be determined to high
  accuracy. This will also reveal differences between the proton and neutron
  polarisabilities and their constituents.
\item[(2)] With a better handle on $\al_{E1}$ and $\be_{M1}$, a combination of
  concurrent unpolarised and polarised measurements can be used to extract the spin
  polarisabilities. Most efficient seems a set of double-polarised experiments
  at $\omega\gtrsim100$~MeV but below the pion-production threshold.
\end{itemize}  
In principle, a multipole-analysis of $4+1$ experiments at different angles
suffices in step (2) to over-determine the $4$ spin-polarisabilities -- if the
data has unprecedentedly high accuracy. $\gamma_{E1E1}$ and $\gamma_{M1M1}$
can to a good degree be extracted uniquely from $\De_x^{circ}$ and
$\De_z^{lin}$, respectively.  However, a larger number of independent data
will be necessary to account for the high complexity of these experiments, and
for the fact that no clear-cut observables exist for the ``mixed''
spin-polarisabilities $\gamma_{E1M2}$ and $\gamma_{M1E2}$. In each, the
accuracy achievable and the observables and kinematics most suited strongly
depend on geometry and acceptance of the experimental setup. We have therefore
made our detailed results available as interactive \emph{Mathematica 7.0}
notebook (email to hgrie@gwu.edu).

In the long run, a multipole-analysis of Compton scattering at fixed energies
from double-polarised, high-accuracy experiments provides a new avenue to
extract the energy-dependence of the six dipole-polarisabilities per
nucleon~\cite{Hi04}. A concerted effort of planned and approved experiments at
$\omega\lesssim200\;\mathrm{MeV}$ is indeed under way: polarised photons on
polarised protons, deuterons and ${}^3$He at TUNL/HI$\gamma$S; tagged protons
at S-DALINAC; polarised photons on polarised protons at MAMI. The unpolarised
experiment on the deuteron at MAXlab over a wide range of energies and angles
is being analysed~\cite{myers}. With at present only 28 (un-polarised)
points for the deuteron in a small energy range of $\omega\in[49;94]$ MeV and
error-bars on the order of $15\%$, high-quality data allow one to zoom in on
the proton-neutron differences and provide first information on the
spin-polarisabilities.  We re-iterate that a publication elaborating
on these findings is forthcoming~\cite{hgds}.


Enlightening insight into the electro-magnetic structure of the nucleon has
already been gained from combining Compton scattering off nucleons and
few-nucleon systems with $\chi$EFT and the (energy-dependent) dynamical
polarisabilities; and a host of activities should add to it in the coming
years, so that we understand the response of the nucleon spin constituents to
external electro-magnetic fields, as parameterised by its
spin-polarisabilities.

\section*{Acknowledgements}
We acknowledge financial support by the National Science Foundation (CAREER
grant PHY-0645498) and US Department of Energy (DE-FG02-95ER-40907).

\end{document}